\documentstyle[12pt]{article}

\newcommand {\e} {\mbox{\rm e}}

\newcounter{eq}
\newcounter{sc}

\def\overleftrightarrow#1{\vbox{\ialign{##\crcr
 $\leftrightarrow$\crcr\noalign{\kern-1pt\nointerlineskip}
 $\hfil\displaystyle{#1}\hfil$\crcr}}}

\newlength{\minitwocolumn}
\begin{document}

\begin{flushright}
DPUR/TH/62\\
December, 2018\\
\end{flushright}
\vspace{20pt}

\pagestyle{empty}
\baselineskip15pt

\begin{center}
{\large\bf Scale Symmetry and Weinberg's No-go Theorem in the Cosmological
Constant Problem
\vskip 1mm }

\vspace{20mm}

Ichiro Oda\footnote{
           E-mail address:\ ioda@sci.u-ryukyu.ac.jp
                  }

\vspace{10mm}
           Department of Physics, Faculty of Science, University of the 
           Ryukyus,\\
           Nishihara, Okinawa 903-0213, Japan\\

\end{center}


\vspace{10mm}
\begin{abstract}

We complete the proof of Weinberg's no-go theorem on the cosmological constant problem in classical gravity 
when the theory has a (global) scale symmetry. Stimulated with this proof, we explore a solution to the 
cosmological constant problem by the help of renormalization group equations. We find that the manifestly 
scale invariant regularization method provides a physically plausible solution to the cosmological constant problem,
in particular, to the issue of radiative instability of the cosmological constant.
 
\end{abstract}

\newpage
\pagestyle{plain}
\pagenumbering{arabic}


\rm
\section{Introduction}

The cosmological constant problem (CCP) has been discussed in uncountably many papers 
since it was realized that radiative corrections in quantum field theory (QFT) give rise to a vacuum 
energy density which is many orders of magnitude greater than that allowed by cosmological observation
\cite{Weinberg, Padilla}. The extremely tiny value of the cosmological constant (CC) at the present epoch, 
which is observed to be around $1 (meV)^4$, appears especially mysterious in considering the fact 
that our universe thus far underwent several phase transitions or symmetry breakings which greatly 
change the value of the CC.

In QFT, symmetry breakings naturally lead to a CC of order $E^4$ where $E$ is the characteristic energy
scale of each symmetry breaking. For instance, the Standard Model (SM) of particle physics informs us
that we have experienced at least two stages of symmetry breakings, those are, the Higgs condensation
$\langle H^\dagger H \rangle \sim M_{EW}^2 \sim (200 GeV)^2$ and the QCD chiral condensation
$\langle \bar q q \rangle \sim \Lambda_{QCD}^3 \sim (200 MeV)^3$. Each symmetry breaking produces 
a huge value of the CC, but the last symmetry breaking must produce a very tiny CC to very high accuracy, 
which is like a magic. This fact might suggest that the CC behaves as if it were completely blind to a huge 
vacuum energy density coming from symmetry breakings.  

In order to account for such a tiny value of the CC, one could envision a few viable scenarios, one of which
is to make use of some symmetry which reduces a large CC to the tiny one or zero. The difficulty with this
symmetry approach is that no appropriate symmetry is known at present which can do such the job. 
However, since the CCP and the gauge hierarchy problem are intrinsically related to energy and mass scales,
it is natural to expect that a (local) conformal symmetry or a (global) scale symmetry might play a role 
to some degree in understanding the two problems. Actually, we will see that the scale symmetry provides
us with a playground for attacking the CCP.

The other popular scenario is to utilize some dynamical mechanism which makes the CC relax to the tiny value
or zero. A natural candidate realizing such a scenario is that some matter field almost "eats up" the large CC, thereby
its small fraction, which the matter field could not eat up, being left behind. This scenario must confront and 
overcome the Weinberg's venerable no-go theorem to provide a plausible solution to the CCP \cite{Weinberg, 
Padilla, Oda1}.

In this article, we wish to pursue the third scenario which is in a sense a compromised scenario between the above 
two scenarios, i.e., symmetry approach and dynamical mechanism. It turns out that it is the Weinberg's no-go theorem 
for a theory with scale invariance that gives us a hint of exploring such a solution to the CCP. 

In general relativity we are familiar with one fact that the presence of the CC makes it impossible for 
a flat Minkowski space-time to become a classical solution to the Einstein equation. This fact has been upgraded to, 
what we call, the Weinberg's no-go theorem, in an attempt to search for a solution to the CCP \cite{Weinberg}. 
The Weinberg's no-go theorem in classical gravity can be stated as follows: General coordinate symmetries, 
which are in general violated by the presence of a fixed background metric, cannot be broken without any fine-tuning 
of the CC in such a way that the translational invariance, which is a subgroup of general coordinate symmetries, 
is exactly preserved. As emphasized by Weinberg \cite{Weinberg}, this situation is unusual from the QFT viewpoint
since given a theory invariant under some gauge group $G$, we would not expect to have to fine-tune the parameters
of the theory to find vacuum solutions which preserve any subgroup $H \subset G$. Incidentally, the Weinberg theorem
in classical gravity is naturally generalized to quantum gravity on the basis of both the BRST invariance and 
the effective action \cite{Nagahama}.

The rest of this paper is organised as follows: In the next section, we review Weinberg's no-go theorem in
classical gravity. In Section 3, we complete a proof of the Weinberg theorem when there is a scale symmetry.
In Section 4, we look for a solution to the CCP by the help of renormalization group equations \cite{Gell-Mann, Callan1, Symanzik}. 
In Section 5, we find that a scale invariant theory provides a solution to the CCP, in particular, to the issue of radiative instability
of the CCP, when renormalizing it in terms of the manifestly scale invariant regularization method. Finally, we conclude 
in Section 6.

\section{Weinberg's theorem in classical gravity}

In this section, we begin by reviewing the Weinberg's no-go theorem in classical gravity \cite{Weinberg}
and point out that there is a loophole in the argument in the presence of scale invariance.

The argument by Weinberg starts with a Lagrangian density ${\cal L}(g_{\mu\nu}, \varphi_A)$ 
which consists of the metric tensor $g_{\mu\nu}$ and generic matter fields $\varphi_A$ 
where the subscript $A$ takes the values $A = 0, 1, 2, \cdots$ and labels different fields 
with suppressed tensor indices. The point in the Weinberg's argument is to assume that the vacuum is 
translationally invariant, by which all fields must be constant in space-time. For such constant fields 
general coordinate symmetries are reduced to a global $GL(4)$ symmetry \cite{Weinberg}-\cite{Nagahama}
\begin{eqnarray}
x^\mu \rightarrow x^{\prime\mu} = (M^{-1})^\mu \,_\nu x^\nu,
\label{GL(4)}
\end{eqnarray}
where $M^\mu \,_\nu$ is a constant $4 \times 4$ matrix satisfying $\det M \neq 0$. Under the $GL(4)$ 
transformation, the constant fields and the Lagrangian density are transformed as 
\begin{eqnarray}
g_{\mu\nu} &\rightarrow& g^\prime_{\mu\nu} = g_{\alpha\beta} M^\alpha \,_\mu M^\beta \,_\nu,
\quad \varphi_A \rightarrow \varphi_A^\prime = D_{AB} (M) \varphi_B, 
\nonumber\\
{\cal L}(g, \varphi_A) &\rightarrow& {\cal L}^\prime (g^\prime, \varphi^\prime_A) = \det M \cdot {\cal L}(g, \varphi_A),
\label{GL(4)-2}
\end{eqnarray}
where $D_{AB}(M)$ is an appropriate representation matrix for the tensor structure of the fields $\varphi_A$.
With $M^\mu \,_\nu = \delta^\mu_\nu + \delta M^\mu \,_\nu$ ($|\delta M^\mu \,_\nu| \ll 1$), the infinitesimal $GL(4)$ 
transformations for the metric tensor and the Lagrangian density read
\begin{eqnarray}
\delta g_{\mu\nu} = \delta M_{\mu\nu} +  \delta M_{\nu\mu}, \qquad
\delta {\cal L} = Tr \delta M \cdot {\cal L},
\label{GL(4)-3}
\end{eqnarray}
where $Tr \delta M \equiv g^{\mu\nu} \delta M_{\mu\nu}$. Note that the latter transformation implies that the 
Lagrangian density indeed transforms as a density under the $GL(4)$ transformation.

Given constant fields, under the infinitesimal $GL(4)$ transformation, the Lagrangian density transforms as
\begin{eqnarray}
\delta {\cal L} = Tr \delta M \cdot {\cal L} = \frac{\partial {\cal L}}{\partial \varphi_A} \delta \varphi_A
+ \frac{\partial {\cal L}}{\partial g_{\mu\nu}} (\delta M_{\mu\nu} +  \delta M_{\nu\mu}).
\label{dL-CG1}
\end{eqnarray}
This relation is used to show that given the matter field equation $\frac{\partial {\cal L}}{\partial \varphi_A} = 0$,
the dependence of ${\cal L}$ on $g_{\mu\nu}$ is too simple to allow a solution to the gravitational field equation
$\frac{\partial {\cal L}}{\partial g_{\mu\nu}} = 0$ unless we pick up the vanishing CC by hand. Choosing the vanishing
CC by hand is interpreted as a fine-tuning of the CC. Incidentally, provided that the matter fields are a scalar field 
$\varphi_A = \varphi$, its $GL(4)$ variation is identically vanishing, $\delta \varphi = 0$, so that the first term 
on the right-hand side (RHS) of Eq. (\ref{dL-CG1}) becomes zero. Even in this case, subsequent arguments are still valid. 

Let us consider two distinct cases separately: one case is that both of the field equations, those are, 
$\frac{\partial {\cal L}}{\partial \varphi_A} = 0$ and $\frac{\partial {\cal L}}{\partial g_{\mu\nu}} = 0$, hold 
independently whereas the other case is that they are not independent and related to each other via a certain relation,
which means the existence of scale invariance in a theory.

For the former case we will first assume $\frac{\partial {\cal L}}{\partial \varphi_A} = 0$ without assuming 
$\frac{\partial {\cal L}}{\partial g_{\mu\nu}} = 0$ for the moment.  Then, Eq. (\ref{dL-CG1}) is simply solved to be
\begin{eqnarray}
{\cal L} = \sqrt{-g} V(\varphi_A),
\label{Matter-sol1}
\end{eqnarray}
where $V(\varphi_A)$ satisfies $\frac{\partial V}{\partial \varphi_A} = 0$ and it is some function depending 
on only matter fields $\varphi_A$. 
As mentioned before, we see that the dependence of ${\cal L}$ on $g_{\mu\nu}$ is too simple to 
allow a solution to the gravitational field equation $\frac{\partial {\cal L}}{\partial g_{\mu\nu}} = 0$ unless $V(\varphi_A)$
is vanishing. Let us recall that choosing $V(\varphi_A) = 0$ by hand corresponds to a fine-tuning of the CC.
In other words, in the case that the metric tensor and matter fields are independent fields, the CCP cannot be
solved except for a fine-tuning of the CC, which is a very well-known result.  So far, so good!

Next, let us turn our attention to the second case where the two field equations are related to each other through
the relation \cite{Weinberg}
\begin{eqnarray}
2 g_{\mu\nu} \frac{\partial {\cal L}}{\partial g_{\mu\nu}} = \sum_A f_A (\varphi) \frac{\partial {\cal L}}{\partial \varphi_A},
\label{M-relation}
\end{eqnarray}
where $f_A (\varphi)$ is a certain function depending on $\varphi_A$. This relation can be rephrased as the existence of
a global symmetry in the theory under consideration
\begin{eqnarray}
\delta_\epsilon g_{\mu\nu} = 2 \epsilon  g_{\mu\nu}, \qquad  \delta_\epsilon \varphi_A = - \epsilon  f_A (\varphi),
\label{Scale-inv}
\end{eqnarray}
where $\epsilon$ is the infinitesimal transformation parameter. By redefining the fields in an appropriate way, 
one can take at least locally \cite{Weinberg}
\begin{eqnarray}
\delta_\epsilon g_{\mu\nu} = 2 \epsilon  g_{\mu\nu}, \qquad  \delta_\epsilon \varphi = - \epsilon,
\qquad  \delta_\epsilon \varphi_a = 0,
\label{Scale-inv2}
\end{eqnarray}
where we have defined $\varphi = \varphi_0$ and $\varphi_a = \varphi_{A \neq 0}$.
Note that this transformation coincides with the conventional scale transformation if we identify $\Phi \equiv e^\varphi$
and $\varphi_a$ with scalar and gauge fields, respectively, and assign mass dimension to the both fields. 
With this identification, it is reasonable to set up a $GL(4)$ transformation, $\delta \varphi = 0$ or 
$\delta \Phi = 0$.  Since we can construct a scale invariant metric
\begin{eqnarray}
\delta_\epsilon ( e^{2 \varphi} g_{\mu\nu} ) = 0,
\label{Scale-inv-metric}
\end{eqnarray}
a scale invariant Lagrangian density can be described as
\begin{eqnarray}
{\cal L} = {\cal L} ( e^{2 \varphi} g_{\mu\nu}, \varphi_a ) \equiv {\cal L} ( \hat g_{\mu\nu}, \varphi_a ),
\label{Scale-inv-Lag}
\end{eqnarray}
where we have introduced the scale invariant metric $\hat g_{\mu\nu} = e^{2 \varphi} g_{\mu\nu}$.

As in the previous case, let us take the variation of ${\cal L}$ under the $GL(4)$ transformation
\begin{eqnarray}
\delta {\cal L} = \frac{\partial {\cal L}}{\partial \varphi_a} \delta \varphi_a
+ \frac{\partial {\cal L}}{\partial \hat g_{\mu\nu}} \delta \hat g_{\mu\nu},
\label{dL-CG2}
\end{eqnarray}
where we have used $\delta \varphi = 0$. Next, following the same line of reasoning as before, we first
impose the matter field equation $\frac{\partial {\cal L}}{\partial \varphi_a} = 0$ without imposing
the gravitational equation $\frac{\partial {\cal L}}{\partial \hat g_{\mu\nu}} = 0$, and then we get the relation
\begin{eqnarray}
\delta {\cal L} = Tr \delta \hat M \cdot {\cal L} = \frac{\partial {\cal L}}{\partial \hat g_{\mu\nu}} 
( \delta \hat M_{\mu\nu} +  \delta \hat M_{\nu\mu}),
\label{dL-CG3}
\end{eqnarray}
where we have defined $\hat M_{\mu\nu} = e^{2 \varphi} M_{\mu\nu}$,  $Tr \delta \hat M = \hat g^{\mu\nu} 
\delta \hat M_{\mu\nu}$ and $\hat g^{\mu\nu} = e^{-2 \varphi} g^{\mu\nu}$. The solution to Eq. (\ref{dL-CG3}) 
reads
\begin{eqnarray}
{\cal L} = \sqrt{- \hat g} V(\varphi_a) = \sqrt{- g} e^{4 \varphi} V(\varphi_a).
\label{Matter-sol2}
\end{eqnarray}

Finally, imposing the gravitational field equation $\frac{\partial {\cal L}}{\partial \hat g_{\mu\nu}} = 0$ requires us 
to take $V(\varphi_a) = 0$ or $e^{4 \varphi} \rightarrow 0$. As before, the former case corresponds to a fine-tuning
of the CC. In order to understand the meaning of the latter case, it is convenient to recall that $\Phi = e^\varphi$
is a scalar field. Then, the limit  $e^{4 \varphi} \rightarrow 0$ is equivalent to the vanishing limit of the scalar field,
$\Phi \rightarrow 0$. Since the scalar field is expected to "eat up" the large CC, this case should be excluded from
our consideration from the physical viewpoint. 

It seems that the above proof completes the Weinberg's no-go theorem in the presence of scale invariance.  
Any proposal for attempting to solve the CCP must confront this theorem and explain how the
proposal escapes its clutches. This is especially true when we assume an approximate translational invariance
and constancy of all fields at small scales. However, it is fortunate that there is a loophole in the proof when 
there is a scale symmetry. In the next section, we shall present a slightly modified but complete proof clarifying
the loophole, and stimulated with this new proof we will explore a new solution to the CCP in the following sections.

\section{New proof of the Weinberg theorem in the presence of scale symmetry}

In the latter part of previous section, we have presented a proof of the Weinberg's no-go theorem in classical
gravity in the presence of scale invariance. However, this proof is incomplete in that it neglects the fact that 
the potential is also constrained by scale symmetry. For instance, in the classically scale invariant $\lambda \phi^4$
theory, the potential $V(\phi)$ satisfies the relation at the classical level
\begin{eqnarray}
T^\mu \, _\mu = 4 V(\phi) - \phi \frac{\partial}{\partial \phi} V(\phi) = 0,
\label{Dim-counting}
\end{eqnarray}
where $T_{\mu\nu}$ is an "improved" stress-energy tensor introduced by Callan et al. \cite{Callan2}.
In fact, it will be found shortly that Eq. (\ref{Dim-counting}) is the key relation for providing a complete proof 
for the theorem.

In this section, we wish to present an alternative and complete proof of the Weinberg's no-go theorem in
classical gravity when a scale symmetry exists. To this end, let us start with a Lagrangian density
${\cal L}(g_{\mu\nu}, \phi_i, \varphi_A)$ with a manifest scale symmetry from the beginning. Here 
$\phi_i$ denote a set of scalar fields and $ \varphi_A$ are matter fields but scalar fields. The infinitesimal
scale transformation is defined in a usual manner as   
\begin{eqnarray}
\delta_\epsilon g_{\mu\nu} = 2 \epsilon  g_{\mu\nu}, \qquad  \delta_\epsilon \phi_i = - \epsilon \phi_i,
\label{Scale-Symm}
\end{eqnarray}
in addition to $\delta_\epsilon \varphi_A$. Of course, as before, it is easy to construct a scale invariant metric
$\hat g_{\mu\nu}$ out of $g_{\mu\nu}$ and a set of scalar fields $\phi_i$ as
\begin{eqnarray}
\hat g_{\mu\nu} = \left( \sum_{i, j} c_{ij} \phi_i \phi_j \right) g_{\mu\nu} \equiv \phi^2 g_{\mu\nu},
\label{Scale-Inv-Metric}
\end{eqnarray}
where $c_{ij}$ is an $n \times n$ constant matrix and we have defined $\phi^2 \equiv \sum_{i, j} c_{ij} \phi_i \phi_j$.

The dynamics is described by the Lagrangian density ${\cal L}(g_{\mu\nu}, \phi_i, \varphi_A)$ at the classical
level and by an effective action $\Gamma$ at the quantum level. For constant fields due to the translational invariance, 
the Lagrangian density and the effective action reduce to the classical potential and the effective potential, respectively. 
For the generality of presentation, in this section, we will consider the effective potential $V(g_{\mu\nu}, \phi_i, \varphi_A)$. 
Then, a scale invariant effective potential takes the form 
\begin{eqnarray}
V(g_{\mu\nu}, \phi_i, \varphi_A) = V(\hat g_{\mu\nu}, \varphi_A).
\label{EP}
\end{eqnarray}
Under the $GL(4)$ transformation, 
\begin{eqnarray}
\delta V = \frac{\partial V}{\partial \varphi_A} \delta \varphi_A + \frac{\partial V}{\partial \hat g_{\mu\nu}} 
\delta \hat g_{\mu\nu},
\label{Var-EP}
\end{eqnarray}
where the scalar fields do not transform under the $GL(4)$ transformation, $\delta \phi_i = 0$.

Next, let us impose the matter field equations $\frac{\partial V}{\partial \phi_i} = \frac{\partial V}{\partial \varphi_A}
= 0$ without doing the gravitational equation $\frac{\partial V}{\partial \hat g_{\mu\nu}} = 0$. Consequently, 
we obtain the relation
\begin{eqnarray}
\delta V = Tr \delta \hat M \cdot V = \frac{\partial V}{\partial \hat g_{\mu\nu}} 
( \delta \hat M_{\mu\nu} +  \delta \hat M_{\nu\mu}).
\label{V-eq}
\end{eqnarray}
The solution to Eq. (\ref{V-eq}) is then of form  
\begin{eqnarray}
V = \sqrt{- \hat g} \hat V_0 (\varphi_A) = \sqrt{- g} \phi^4 \hat V_0 (\varphi_A) 
\equiv \sqrt{- g} V_0 (\phi_i, \varphi_A),
\label{Matter-sol3}
\end{eqnarray}
where $V_0 (\phi_i, \varphi_A) \equiv \phi^4 \hat V_0 (\varphi_A)$.  Since the constant fields $\varphi_A$, 
which do not include scalar fields, do not play a special role any more, let us put $\varphi_A = 0$ for simplicity. 
Accordingly, we arrive at the solution
\begin{eqnarray}
V = \sqrt{- g} V_0 (\phi_i),
\label{Matter-sol4}
\end{eqnarray}
where $V_0 (\phi_i) \equiv V_0 (\phi_i, \varphi_A = 0)$ and $\phi_i$ are constant scalar fields satisfying 
$\frac{\partial V_0}{\partial \phi_i} = 0$. 

Finally, it seems that imposing the gravitational field equation $\frac{\partial V}{\partial g_{\mu\nu}} = 0$
leads to the previous result, $V_0 (\phi_i) = 0$, which corresponds to a fine-tuning of the CC. However,
there is a loophole at this point. At the classical level, since there is no trace anomaly, as in Eq. (\ref{Dim-counting})
the scale invariant potential $V_0 (\phi_i)$ must satisfy the relation
\begin{eqnarray}
4 V_0 (\phi_i) - \sum_j \phi_j \frac{\partial}{\partial \phi_j} V_0 (\phi_i) = 0.
\label{Class-RGE}
\end{eqnarray}
Now $V_0 (\phi_i)$ also satisfies the field equation $\frac{\partial V_0}{\partial \phi_i} = 0$, so the relation
(\ref{Class-RGE}) requires us to take
\begin{eqnarray}
V_0 (\phi_i) = 0.
\label{Zero-Pot}
\end{eqnarray}
Hence, at least at the classical level, the potential $V_0 (\phi_i)$ is automatically zero, which means that we do not
have to fine-tune the CC by hand.
This new proof not only demonstrates that the Weinberg's no-go theorem does not hold at least at the classical level 
when a theory possesses a scale symmetry, but also strongly suggests that a scale symmetry could play an important 
role for the CCP since the Weinberg's no-go theorem has been regarded as closing off all hopes in this direction 
thus far. However, the point is that a scale symmetry is explicitly broken by trace anomaly at the quantum level, 
so the above conclusion is limited to hold only at the classical level.

\section{A solution to the cosmological constant problem}

We would like to explore a solution to the cosmological constant problem (CCP).  As pointed out in the last section,
the point is to deal with the breaking of scale symmetry at the quantum level which ruins nice properties of a classical
scale symmetry. In this section, we use the semiclassical approach where only matter fields are treated as quantum fields 
whereas the gravitational field is regarded as a classical field, i.e., a fixed classical backgound. It is worthwhile to 
stress that the semiclassical approach is physically plausible in finding a solution to the CCP since the CCP stems from 
a clash between particle physics which sources the vacuum energy density via quantum effects and gravity responding to 
its vacuum energy density classically. Moreover, it is sufficient to take a Minkowski background, $g_{\mu\nu}= \eta_{\mu\nu}$ 
for understanding the essential point of the CCP in the framework of the semiclassical approach. 

We also adopt the viewpoint of the effective field theory: even if a scale symmetry at some high energy is broken spontaneously, 
a new scale symmetry at the lower energy reappears after integrating out massive states made by the scale symmetry breaking 
at the high energy. This viewpoint is new and has been recently advocated as the multi-step spontaneous breaking of scale 
symmetry by Kugo \cite{Kugo}, so let us explain it briefly: First, suppose that our world has no dimensional parameters, thereby 
realizing a classical scale invariance. Then, also suppose that the total physical system is composed of three kinds of scale invariant 
potentials
\begin{eqnarray}
V (\phi) = V_1 (\sigma) + V_2 (\sigma, h) + V_3 (\sigma, h, \varphi),
\label{3-pot}
\end{eqnarray}
where the vacuum expectation value (VEV) of each field is assumed to have
\begin{eqnarray}
\langle \sigma \rangle = M_{Pl}, \quad \langle h \rangle = M_{EW}, \quad \langle \varphi \rangle = \Lambda_{QCD},
\label{3-pot VEV}
\end{eqnarray}
where $M_{Pl} \simeq 2.4 \times 10^{18} GeV$ denotes the Planck mass scale.
Because of the scale invariance, the vacuum energy at any stationary points, $\langle \phi \rangle = \phi_0$, is
vanishing, $V (\phi_0) = 0$. The important point in the multi-step spontaneous breaking is that this holds at
every stage of spontaneous symmetry breakings. Concretely explaining, in the total potential $V (\phi)$, we can retain
only $V_1 (\sigma)$ in discussing the physics at the scale $M_{Pl}$ since the fields $h$ and $\varphi$ are expected
to have VEVs of the lower scales. Then, the scale invariance ensures $V_1 (\sigma_0) = 0$. When we discuss the
next stage of spontaneous symmetry breaking of scale symmetry at the scale $M_{EW}$, we take the potential
$V_1 (\sigma) + V_2 (\sigma, h)$ and can conclude $V_1 (\sigma_0^\prime) + V_2 (\sigma_0^\prime, h_0) = 0$.
Similarly, at the third stage of symmetry breaking at the scale $\Lambda_{QCD}$, the potential must be chosen to be
$V_1 (\sigma) + V_2 (\sigma, h) + V_3 (\sigma, h, \varphi)$, and we can then conclude $V_1 (\sigma_0^{''}) 
+ V_2 (\sigma_0^{''}, h_0^\prime) + V_3 (\sigma_0^{''}, h_0^\prime, \varphi_0) = 0$. In this way, 
the vanishing CC can be realized at each stage of symmetry breaking of scale symmetry since the scale symmetry is
valid at every energy scale of spontaneous symmetry breaking.

In order to consider quantum aspects of the theory, it is useful to start with the conventional dimensional regularization 
(DR) of renormalizable QFTs and work with the renormalization group equation (RGE) for the effective potential 
$V(\phi)$ whose form is given by \cite{Gell-Mann, Callan1, Symanzik}
\begin{eqnarray}
\left[ \mu \frac{\partial}{\partial \mu} + \sum_a \beta_a (\lambda) \frac{\partial}{\partial \lambda_a}
+ \sum_i \gamma_i (\lambda) \phi_i \frac{\partial}{\partial \phi_i} \right] V(\phi) = 0,
\label{RGE 1}
\end{eqnarray}
where $\mu$ is the renormalization mass, and $\beta_a (\lambda)$ and $\gamma_i (\lambda)$ are the beta function 
and the anomalous dimension, respectively. The solution to this RGE is well known and reads \cite{KMN}
\begin{eqnarray}
V(\phi_i, \lambda_a; \mu^2) = V(\bar \phi_i (t), \bar \lambda_a (t) ; e^{2t} \mu^2),
\label{RGE-sol}
\end{eqnarray}
where $\bar \phi_i (t)$ and $\bar \lambda_a (t)$ are running parameters whose $t$-dependence is determined by
\begin{eqnarray}
\frac{d \bar \lambda_a (t)}{d t} = \beta (\bar \lambda_a (t)), \quad 
\frac{d \bar \phi_i (t)}{d t} = - \gamma_\phi (\bar \lambda_a (t)) \bar \phi_i (t),
\label{RGE-para}
\end{eqnarray}
with intial conditions $\bar \lambda_a (0) = \lambda_a$ and $\bar \phi_i (0) = \phi_i$.

In the case at hand, since the effective potential $V(\phi)$ is a homogeneous function, having only a dimensional
mass parameter $\mu$, we have the relation at the quantum level
\begin{eqnarray}
\left( \mu \frac{\partial}{\partial \mu} + \sum_i \phi_i \frac{\partial}{\partial \phi_i} \right) V(\phi) = 4 V(\phi).
\label{Q-Dim-counting}
\end{eqnarray}
Eliminating the term $\mu \frac{\partial V}{\partial \mu}$ from Eqs. (\ref{RGE 1}) and (\ref{Q-Dim-counting}),
we obtain
\begin{eqnarray}
\left[ \sum_i \left( 1 - \gamma_i (\lambda) \right) \phi_i \frac{\partial}{\partial \phi_i} 
- \sum_a \beta_a (\lambda) \frac{\partial}{\partial \lambda_a} \right] V(\phi) = 4 V(\phi).
\label{Rev-RGE 1}
\end{eqnarray}
At stationary points $\langle \phi_i \rangle = \phi_i ^{(0)}$ such that 
\begin{eqnarray}
\left. \frac{\partial V(\phi)}{\partial \phi_i} \right|_{\langle \phi_i \rangle = \phi_i ^{(0)}}  = 0,
\label{Extr-pot}
\end{eqnarray}
Eq. (\ref{Rev-RGE 1}) reduces to the form
\begin{eqnarray}
V(\phi ^{(0)}) = - \frac{1}{4} \sum_a \beta_a (\lambda) \left. \frac{\partial V(\phi)}{\partial \lambda_a} 
\right|_{\langle \phi_i \rangle = \phi_i ^{(0)}}.
\label{Stat-pot}
\end{eqnarray}
This relation informs us that the anomalous dimension $\gamma_i (\lambda)$ is irrelevant to the CCP
while the beta function $\beta_a (\lambda)$ makes a contribution to the CC. 

An obvious possibility for the vanishing CC is that all the coupling constants $\lambda_a$ are
attracted to the infrared fixed points $\lambda_a ^{(IR)}$
\begin{eqnarray}
\beta_a (\lambda^{(IR)}) = 0,
\label{IR-fixed point}
\end{eqnarray}
where 
\begin{eqnarray}
\lambda_a ^{(IR)} \equiv \lim_{t \rightarrow - \infty} \bar \lambda_a (t).
\label{Def-IR-fixed point}
\end{eqnarray}
However, since our world is not scale invariant at low energies, this possibility is not physically appealing
and should be thrown away. Another more interesting possibility pointed out in Ref. \cite{Kugo} is that
the potential $V(\phi^{(0)})$ at the stationary points is vanishing at any scale $\mu$ even before reaching
the infrared limit $\mu \rightarrow 0$ or $t \rightarrow - \infty$.\footnote{$\mu$ and $t$ are related 
by the relation $\mu = \mu_0 \e^t$ where $\mu_0$ is a certain constant having mass dimension.}  
However, it seems that this possibility might be realized only when the scale invariant theories could generate
a non-zero scale spontaneously. 
  
Thus, in order to solve the CCP on the basis of scale invariance, we should explore alternative possibilities.
One promising possibility is to make use of spontaneous symmetry breakdown of scale symmetry which
is derived via the manifestly scale invariant dimensional regularization (SD) \cite{Englert}-\cite{Tamarit}.
In the standard method for evaluating radiative corrections a classical scale symmetry is explicitly violated 
through the regularization procedure involving a subtraction scale such as the renormalization scale $\mu$
in the dimensional regularization (DR) and the ultraviolet scale $\Lambda$ in the cut-off regularization. In the
SD procedure, the renormalization scale $\mu$ in the DR is replaced with a scalar
field which acquires a vacuum expectation value (VEV) after spontaneous symmetry breaking of scale
symmetry. In the whole process of renormalization, the scale symmetry is maintained and we have a
Goldstone mode called "${\it{dilaton}}$" which is exactly massless and remains a flat direction of an
effective potential. The point in the SD scheme is that all masses are generated from a VEV of the dilaton
and the requirement that the bare coupling constants are independent of a parameter relating the renormalization
scale to the dilaton naturally leads to the running of renormalized coupling constants which is the same as that
obtained via the conventional DR method.

Of course, there are some disadvantages in the SD procedure. First of all, an exact scale invariance at the 
quantum level is preserved at the cost of non-renormalizability and the theory therefore makes sense only
as an effective field theory. However, this issue may not be problematic when coupling to general relativity
since general relativity is in itself non-renormalizable.\footnote{We have recently derived general relativity from
conformal gravity \cite{Oda2}. In this case, since conformal gravity is renormalizable, the non-renormalizabilty becomes
a problem.} Secondly, it is not clear how to define the unbroken phase of scale symmetry in the SD scheme. 
In the usual QFT, the unbroken phase, $\langle \sigma \rangle = 0$, is smoothly connected with the broken phase, 
$\langle \sigma \rangle \neq 0$ whereas in the SD scheme, the VEV of the dilaton appears in the denominator 
of various equations so it is difficult to take the limit, $\langle \sigma \rangle \rightarrow 0$. This situation
makes it difficult to understand a relation between the unbroken phase and the broken phase in the SD method.
Finally, it is not clear either whether or not the flat direction in the potential is maintained against quantum corrections. 
This problem is very important in considering the CCP within the present formulation, so it will be discussed in detail 
in the next section.
  
In the SD method for the scale invariant theories, the RGE for the effective potential in general takes the form
\begin{eqnarray}
\left[ z \frac{\partial}{\partial z} + \sum_a \beta_a (\lambda) \frac{\partial}{\partial \lambda_a}
+ \sum_A \beta_A (\lambda) \frac{\partial}{\partial \lambda_A}
+ \sum_i \gamma_i (\lambda) \phi_i \frac{\partial}{\partial \phi_i} \right] V(\phi) = 0,
\label{RGE 2}
\end{eqnarray}
where $z$ is a parameter relating the renormalization scale to the dilaton, and $\lambda_A$ are the coupling constants 
generated by the non-renormalizability of the theory.  Since we have not introduced
the independent renormalization mass scale and the scale invariance is maintained even at the quantum level, the dimension 
counting equation is the same as that of the classical case:
\begin{eqnarray}
\sum_i \phi_i \frac{\partial}{\partial \phi_i} V(\phi) = 4 V(\phi).
\label{Q-dim-counting}
\end{eqnarray}
Thus, at any stationary points, $\langle \phi_i \rangle = \phi_i ^{(0)}$,  we have
\begin{eqnarray}
V(\phi^{(0)}) = 0,
\label{Q-zero CC}
\end{eqnarray}
which implies that the CC is vanishing even at the quantum level in addition to the classical level due to the exact scale
invariance. Incidentally, at stationary points, Eq. (\ref{RGE 2}) is reduced to the form
\begin{eqnarray}
\left[ z \frac{\partial}{\partial z} + \sum_a \beta_a (\lambda) \frac{\partial}{\partial \lambda_a}
+ \sum_A \beta_A (\lambda) \frac{\partial}{\partial \lambda_A} \right] V(\phi^{(0)}) = 0,
\label{RGE 2 at Stat}
\end{eqnarray}
which must hold at the quantum level in an exact manner.

\section{A simple scale invariant model with two scalar fields}

The aim of this section is to consider the issue of a flat direction in the scale invariant theories. This issue has been
already investigated in \cite{Shaposhnikov}, and we would like to examine it again by using a more detailed model
developed in \cite{Ghilencea1} since the problem of the flat direction is closely related to the fine-tuning problem 
in the CCP.

As a concrete model of the scale invariant theories, let us work with the simple and familiar Lagrangian density with
two real scalar fields \cite{Ghilencea1}
\begin{eqnarray}
{\cal {L}} = - \frac{1}{2} \partial_\mu \phi \partial^\mu \phi - \frac{1}{2} \partial_\mu \sigma \partial^\mu \sigma
- V(\phi, \sigma),
\label{Two-scalar model}
\end{eqnarray}
where we call the scalar fields $\phi$ and $\sigma$ the Higgs field and the dilaton, respectively, and
the scale invariant potential $V(\phi, \sigma)$ is of form 
\begin{eqnarray}
V(\phi, \sigma) = \frac{\lambda_\phi}{4} \phi^4 + \frac{\lambda_m}{2} \phi^2 \sigma^2 
+  \frac{\lambda_\sigma}{4} \sigma^4,
\label{Scale-inv pot}
\end{eqnarray}
where $\lambda_\phi, \lambda_m$ and $\lambda_\sigma$ are dimensionless coupling constants. Note that this Lagrangian
density has a discrete symmetry: It is invariant under the operation $\phi \rightarrow - \phi, 
\sigma \rightarrow - \sigma$. 

The stationary conditions for the classical potential, $V_\phi \equiv \partial_\phi V = 0$ and $V_\sigma \equiv 
\partial_\sigma V = 0$, yield two equations:
\begin{eqnarray}
\phi (\lambda_\phi \phi^2 + \lambda_m \sigma^2) = 0, \quad  
\sigma (\lambda_m \phi^2 + \lambda_\sigma \sigma^2) = 0.
\label{Stat-cond}
\end{eqnarray}
We therefore have two configurations with the minimum energy, one of which is a symmetric ground state, 
$\langle \sigma \rangle = \langle \phi \rangle = 0$ where both the dilaton and the Higgs boson are massless. 
Hereafter, we will ignore this trivial configuration. The second configuration, which is more interesting, is that   
of spontaneous symmetry breakdown of scale symmetry when $\langle \sigma \rangle \neq 0$:
\begin{eqnarray}
\frac{\langle \phi \rangle^2}{\langle \sigma \rangle^2 } = - \frac{\lambda_m}{\lambda_\phi},
\quad \lambda_m^2 = \lambda_\phi \lambda_\sigma,
\label{SSB of scale symm}
\end{eqnarray}
where $\lambda_\phi > 0$ and $\lambda_\sigma > 0$ owing to the stability of the potential while $\lambda_m < 0$ 
from Eq. (\ref{SSB of scale symm}). In this case, the potential contains two flat directions, $\langle \phi \rangle
= \pm \sqrt{- \frac{\lambda_m}{\lambda_\phi}} \langle \sigma \rangle$, and the vacuum is degenerate.  Let us
notice that the spontaneous symmetry breakdown of scale symmetry also triggers the electroweak symmetry 
breakdown at the tree level with a vanishing CC, and all particle masses are generated by the VEV of the dilaton, 
$\langle \sigma \rangle$.

In the manifestly scale invariant dimensional regularization (SD) procedure, the subtraction scale is a generic
function of the dilaton and the Higgs field, $\mu = \mu(\phi, \sigma)$. But the requirement that quantum interactions
between $\phi$ and $\sigma$ switch off in the classical decoupling limit, $\lambda_m \rightarrow 0$, leads to
the condition that the subtraction scale function $\mu(\phi, \sigma)$ is a function depending on only the dilaton
\begin{eqnarray}
\mu(\phi, \sigma) = \mu(\sigma) = z \sigma,
\label{mu}
\end{eqnarray}
where $z$ is an arbitrary dimensionless parameter.

Then, by evaluating the determinant of quadratic terms in the action, the one-loop effective potential is 
obtained via the SD method in $d = 4 - 2 \epsilon$
\begin{eqnarray}
U = \mu(\sigma)^{2 \epsilon} \left\{ V(\phi, \sigma) - \frac{1}{64 \pi^2} \left[ \sum_{s = \phi, \sigma} 
M_s ^4 \left( \frac{1}{\epsilon} - \log \frac{M_s ^2}{\kappa \mu(\sigma)^2} \right) 
+ \frac{4 Tr (M^2 N)}{\mu(\sigma)^2} \right] \right\},
\label{1-loop EP}
\end{eqnarray}
where $\kappa \equiv 4 \pi \e^{\frac{3}{2} - \gamma_E}$ ($\gamma_E \approx 0.5772$ is the 
Euler-Mascheroni constant), $M_s ^2$ is the eigenvalue of the matrix $(M^2)_{\alpha\beta} \equiv 
\partial_\alpha \partial_\beta V \equiv V_{\alpha\beta}$, and   
\begin{eqnarray}
N _{\alpha\beta} \equiv \mu ( \mu_\alpha V_\beta + \mu_\beta V_\alpha ) - \mu_\alpha \mu_\beta V,
\label{N}
\end{eqnarray}
with $\mu_\alpha \equiv \partial_\alpha \mu$ and $V_\alpha \equiv \partial_\alpha V$ ($\alpha, \beta
= \phi, \sigma$).

The counter-terms removing the divergences associated with the simple pole $\frac{1}{\epsilon}$
are given by
\begin{eqnarray}
U_{ct} = \mu(\sigma)^{2 \epsilon} \left[ a_1 \phi^4 \left( \frac{1}{\bar \epsilon} + c_1 \right) 
+ a_2 \phi^2 \sigma^2 \left( \frac{1}{\bar \epsilon} + c_2 \right) 
+ a_3 \sigma^4 \left( \frac{1}{\bar \epsilon} + c_3 \right) \right],
\label{Counter}
\end{eqnarray}
where we have defined $\frac{1}{\bar \epsilon} = \frac{1}{\epsilon} + \log(4 \pi) - \gamma_E$.
The standard $\overline{MS}$ corresponds to the case of $c_1 = c_2 = c_3 = 0$.  Here the constants
$a_i (i = 1, 2, 3)$ turn out to be given by
\begin{eqnarray}
a_1 &=& \frac{1}{64 \pi^2} \left( 9 \lambda_\phi + \frac{\lambda_m ^2}{\lambda_\phi}  \right) \lambda_\phi,
\nonumber\\
a_2 &=& \frac{1}{32 \pi^2} \left( 3 \lambda_\phi + 4 \lambda_m + 3 \lambda_\sigma  \right) \lambda_m,
\nonumber\\
a_3 &=& \frac{1}{64 \pi^2} \left( 9 \lambda_\sigma + \frac{\lambda_m ^2}{\lambda_\sigma}  \right) \lambda_\sigma.
\label{Constants-a}
\end{eqnarray}
With these constants, one finds the finite one-loop effective potential 
\begin{eqnarray}
U_{1-loop} &=& U + U_{ct}      \nonumber\\
&=& V(\phi, \sigma) + \frac{1}{64 \pi^2} \left[ \sum_{s = \phi, \sigma} 
M_s ^4 \left( \log \frac{M_s ^2}{z^2 \sigma^2} - \frac{3}{2} \right) + \Delta U \right],
\label{Fin-1-loop EP}
\end{eqnarray}
where $\Delta U$ is defined as
\begin{eqnarray}
\Delta U &=& \left( 9 \lambda_\phi + \frac{\lambda_m ^2}{\lambda_\phi}  \right) c_1 \lambda_\phi \phi^4
+ 2 \left( 3 \lambda_\phi + 4 \lambda_m + 3 \lambda_\sigma  \right) c_2 \lambda_m \phi^2 \sigma^2
+ \left( 9 \lambda_\sigma + \frac{\lambda_m ^2}{\lambda_\sigma}  \right) c_3 \lambda_\sigma \sigma^4
\nonumber\\
&+& \lambda_\phi \lambda_m \frac{\phi^6}{\sigma^2} - \left( 16 \lambda_\phi \lambda_m 
+ 6 \lambda_m^2 - 3 \lambda_\phi \lambda_\sigma \right) \phi^4
- \left( 16 \lambda_m + 25 \lambda_\sigma \right) \lambda_m \phi^2 \sigma^2
- 21 \lambda_\sigma^2 \sigma^4.
\label{Delta U}
\end{eqnarray}
Eq. (\ref{Fin-1-loop EP}) is a scale invariant one-loop result where the terms except $\Delta U$ coincides with
the Coleman-Weinberg potential \cite{Coleman} if $\mu$ is a constant while $\Delta U$ is a new term induced by 
scale invariance. Note that the appearance of the $\frac{\phi^6}{\sigma^2}$ term expresses the non-renormalizability
of the SD method.  

At this stage, let us assume that 
\begin{eqnarray}
\lambda_\sigma \ll |\lambda_m| \ll \lambda_\phi,
\label{Pheno-ansatz}
\end{eqnarray}
which is true for phenomenological applications. To enforce this hierarchy, it is convenient to introduce a
small parameter $\varepsilon \equiv \left( \frac{M_{EW}}{M_{Pl}} \right)^2 \simeq 
\left( \frac{10^2 GeV}{10^{18} GeV} \right)^2 = 10^{-32} \ll 1$, and regard the magnitude of coupling 
constants as $\lambda_\phi = {\cal{O}}(1)$, $\lambda_m = {\cal{O}}(\varepsilon)$, and 
$\lambda_\sigma = {\cal{O}}(\varepsilon^2)$.    With these assumptions and the classical relation
$\frac{\langle \phi \rangle^2}{\langle \sigma \rangle^2 } \sim  \frac{\lambda_m}{\lambda_\phi}$ in
Eq. (\ref{SSB of scale symm}), one can approximate $\Delta U$ as
\begin{eqnarray}
\Delta U \simeq  9 \lambda_\phi^2 c_1 \phi^4 + 6 \lambda_\phi \lambda_m c_2 \phi^2 \sigma^2
+ \lambda_m ^2 c_3 \sigma^4.
\label{Appro-Delta U}
\end{eqnarray}
Moreover, with these assumptions, the square of two eigenvalues $M_s ^2$ is given by
\begin{eqnarray}
M_1^2 \simeq  3 \lambda_\phi \phi^2 + \lambda_m \sigma^2 + {\cal{O}}(\lambda_m^2), \quad
M_2^2 \simeq  \lambda_m \phi^2 + {\cal{O}}(\lambda_m^2).
\label{M-square}
\end{eqnarray}
Then, the requirement that the classical flat directions, $\langle \phi \rangle = 
\pm \sqrt{- \frac{\lambda_m}{\lambda_\phi}} \langle \sigma \rangle$, are not lifted by quantum effects,
which is equivalent to $\partial_\phi U_{1-loop} = \partial_\sigma U_{1-loop} = 0$ at the classical flat directions,
gives us the relations
\begin{eqnarray}
c_2 &=& 3 c_1 - 2 + 2 \log \frac{-2 \lambda_m}{z^2},   \nonumber\\
c_3 &=& 9 c_1 - 6 + 8 \log \frac{-2 \lambda_m}{z^2} - 16 \pi^2 \left( \frac{\lambda_\sigma}{\lambda_m ^2}
- \frac{1}{\lambda_\phi} \right).
\label{c-relations}
\end{eqnarray}
With these relations (\ref{c-relations}), it is easy to check that the one-loop effective potential is vanishing 
at the classical flat directions, which means that the CC is zero at the one-loop level in addition to the tree level.

Here it is worth reflecting what we have done thus far in this section and its implications for the CCP. 
It is well known that a flat direction at the tree level is not generally maintained at the quantum level 
if there is no symmetry for protecting the flat direction. In the process of renormalization the finite parts of 
the three coupling constants, $\lambda_\phi, \lambda_m$ and $\lambda_\sigma$, are fixed by renormalization 
conditions. To put differently, what we have shown above is that under the assumption (\ref{Pheno-ansatz}) 
the flat direction is still preserved at the one-loop level by taking two renormalization conditions. Even if we have 
shown this procedure explicitly at the one-loop level, there is no obstruction to repeat this procedure at all orders 
of perturbation theory. In QFT, imposing renomalization conditions at each stage of perturbation theory is the
standard procedure, so we have no problem with this fine-tuning. Of course, the relation $\lambda_m^2 = \lambda_\phi
\lambda_\sigma$ at the tree level in Eq. (\ref{SSB of scale symm}) is a true fine-tuning and might be interpreted
as an incarnation of the CCP. However, in order to construct a phenomenologically viable model, a scale symmetry 
must be broken spontaneously at low energies such that the dilaton has a non-vanishing VEV. This situation demands
that the classical potential should have flat directons. In this sense, the relation $\lambda_m^2 = \lambda_\phi \lambda_\sigma$ 
is not a simple fine-tuning but the requirement for making a viable perturbation theory from the scale invariant theories.

However, it should be stressed that the CCP is not a problem of a mere fine-tuning. In some respects, the biggest problem 
of the CCP is that it is rarely stated properly. To solve it, we had better make clear what the problem really is. 
In order to account for the essence of the CCP, as the simplest example, let us consider a real scalar field of mass 
$m$ with the $\lambda \phi^4$ self-coupling, which is minimally coupled to the classical gravity (Gravity is a purely 
classical field merely serving the purpose of detecting vacuum energy):
\begin{eqnarray}
S = \int d^4 x \sqrt{-g} \left[ \frac{M_{Pl}^2}{2} R - \Lambda_b -  \frac{1}{2} g^{\mu\nu} \partial_\mu \phi 
\partial_\nu \phi - \frac{m^2}{2} \phi^2 - \frac{\lambda}{4 !} \phi^4 \right], 
\label{Gravity}
\end{eqnarray}
where $R$ is the scalar curvature\footnote{We will follow the conventions and notation by Misner et al. \cite{MTW}.} 
and $\Lambda_b$ is a bare CC which is divergent. Using the standard dimensional regularization, it is straightforward 
to calculate the one-loop contribution of the scalar field to the CC:
\begin{eqnarray}
U_1 &=& \frac{i}{2} Tr \left[ \log \left( -i \frac{\delta^2 S}{\delta \phi^2} \right) \right] 
\nonumber\\
&=& \frac{1}{2} \int \frac{d^4 p_E}{(2 \pi)^4} \log ( p_E^2 + m^2 )
\nonumber\\
&=& - \frac{m^4}{64 \pi^2} \left[ \frac{1}{\epsilon} + \log \left( \frac{\mu^2}{m^2} \right) + \cdots \right],
\label{1-loop U}
\end{eqnarray}
where the ellipsis denotes the remaining finite parts. The divergence associated with the simple pole requires us
to take the bare CC which depends on an arbitrary subtraction scale $M$:
\begin{eqnarray}
\Lambda_b^{1-loop} = \frac{m^4}{64 \pi^2} \left[ \frac{1}{\epsilon} + \log \left( \frac{\mu^2}{M^2} \right) \right].
\label{1-loop bare CC}
\end{eqnarray}
Then, the one-loop renormalized CC is given by
\begin{eqnarray}
\Lambda_{ren}^{1-loop} = \Lambda_b^{1-loop} + U_1 = \frac{m^4}{64 \pi^2} \left[ \log \left( \frac{m^2}{M^2} \right) 
- \cdots \right].
\label{1-loop ren-CC}
\end{eqnarray}
This 1-loop renormalized CC is finite, but depends on the arbitrary scale $M$, so we cannot have a concrete prediction. 
According to QFT, what we need to do is to replace it with the measured value, not predict the value theoretically. 
Once this is done, one can go on and make predictions about all the physical quantities which are not ultraviolet (UV)
sensitive. 

Cosmological observation requires us to take
\begin{eqnarray}
\Lambda_{ren}^{1-loop} \sim 1 (meV)^4.
\label{Cos-obs}
\end{eqnarray}
If the particle mass $m$ is around $1 TeV$,  
\begin{eqnarray}
U_1 \sim 1 (TeV)^4 = 10^{60} (meV)^4,
\label{U-1}
\end{eqnarray}
Eq. (\ref{1-loop ren-CC}), together with Eqs.  (\ref{Cos-obs}) and (\ref{U-1}), suggests that the finite 
contribution to the 1-loop renormalized CC is cancelled to an accuracy of one part in 
$10^{60}$ between $U_1$ and $\Lambda_b^{1-loop}$. (This big fine tuning is sometimes called the CCP as well.)
Following the lore of QFT, at this stage of the argument, we have no issue with this fine-tuning. However,
the problem arises when we go to higher loops. Namely, at the $n$-loop level, the effective potential $U_n$ is proportional
to be $\lambda^{n-1} m^4$, where the coupling constant $\lambda$ is about ${\cal{O}}(0.1)$ for the SM Higgs.
Thus, at the full quantum level, the renormalized CC must satisfy the relation
\begin{eqnarray}
1 (meV)^4 = \Lambda_{ren} = \Lambda_b + U_1 + U_2 + U_3 + \cdots.
\label{n-loop ren-CC}
\end{eqnarray}
For perturbative theories, the cancellation at the 1-loop level is spoilt, so at the 2-loop level we must retune the finite 
contribution in the bare CC term to the same degree of accuracy owing to ${\cal{O}} (U_1) \simeq  {\cal{O}} (U_2) 
\simeq  {\cal{O}} (U_3) \simeq \cdots$, compared with $\Lambda_{ren} =1 (meV)^4$. 

In other words, at each successive order in perturbation theory, we are required to fine-tune the CC to extreme accuracy!  
This issue is called "${\it radiative \, instability}$" in the CCP, i.e., the need to repeatedly fine tune greatly whenever 
the higher loop corrections are included, which is the essential property of the CCP.  What this is telling us is that the CC is 
very sensitive to the details of UV physics which we are ignorant in the effective theory \cite{Kaloper1}-\cite{Oda6}.  

On the other hand, in the scale invariant theories under consideration, quantum corrections to the CC are so small that 
we are free from the issue of radiative instability and we are happy to tolerate one fine-tuning, that is, the classical fine-tuning 
$\lambda_m^2 = \lambda_\phi \lambda_\sigma$, but not order by order retunings.  Also note that despite the existence
of an infinite number of operators in the SD procedure which ultimately ensure scale invariance of the quantum theory,
they are all supressed by the VEV of the dilaton, i.e., the Planck mass. Consequently, such an infinite number of new operators
coming from the non-renormalizability do not change the one-loop result so much. Related to this fact, let us recall that
an essential point of the CCP is that the vacuum energy associated with various symmetry breakings such as the Higgs
condensation and the QCD chiral condensation must cancel neatly, leaving only its very tiny value behind. It is obvious that
the energy scale of such symmetry breakings is far below the Planck scale, so that it is sufficient to take account of 
the low energy physics and classical gravity. Hence, the large field regime such as $\langle h \rangle \sim M_{Pl}$ (where
$h$ is the Higgs field) and quantum gravity are irrelevant to the resolution of the CCP.

\section{Discussion}

In this article, we have explored a solution to the cosmological constant problem (CCP) by the help of renormalization group
equations. The motivation behind this study is that the complete proof of the Weinberg's no-go theorem in the presence
of scale symmetry suggests that the scale symmetry might play an important role for providing for a solution to the CCP.
 
In applying the scale symmetry for the CCP, we always encounter the problem that the scale symmetry is broken by
trace anomaly at the quantum level, so a nice property that scale invariance forbids the cosmological constant (CC) to
exist in the classical action is not valid any longer in the quantum regime.  In order to overcome this impasse, we have
used a manifestly scale invariant dimensional regularization (SD) method where the scale symmetry is spontaneously
broken owing to the nonzero VEV of the dilaton field, for which the classical potential must have flat directions.
The point is that flat directions are not generally preserved but lifted by quantum corrections. However, we have shown 
that in the phenomenologically viable situation where the dilaton and the Higgs field have VEVs of the Planck scale and
the electroweak scale, respectively, the flat directions are maintained by taking appropriate renormalization conditions.
Compared with the conventional approaches for the CCP, radiative corrections are very mild in the present theory, 
so our solution to the CCP is free from the issue of radiative instability of the CCP. Since we accept to tolerate 
one fine-tuning, that is, the classical fine-tuning $\lambda_m^2 = \lambda_\phi \lambda_\sigma$, but not order by order 
retunings, the present approach sheds light on the resolution to the CCP, in particular, the problem of the radiative
instability of the CCP. 

There are still several questions to be clarified in future. Here is only a partial list of them. First of all, it is absolutely
unclear how to give a very tiny value to the CC since the present formulation naturally produces the vanishing CC.
Secondly, the present theory is not renormalizable so it makes sense only as an effective theory. We wish to understand 
the high energy theory of this theory. Finally, in this article, we have never touched another important hierarchy problem, 
that is, the gauge hierarchy problem.  We conjecture that scale symmetry also plays a critical role in accounting for this problem.
But it is not clear at present how the stability of the electroweak scale against radiative corrections is achieved.  
We wish to return these important problems in future.

\begin{flushleft}
{\bf Acknowledgements}
\end{flushleft}
The work of I. O.  was supported by JSPS KAKENHI Grant Number 16K05327.


\end{document}